\documentclass[a4paper,11pt]{article}
\pdfoutput=1 

\usepackage{jinstpub} 

\title{\boldmath Development and characterization of a large area silicon pad array for an electromagnetic calorimeter}

\author[a,c,1]{Sourav Mukhopadhyay,\note{Corresponding author.}}
\author[a,c]{Vinay B. Chandratre,}
\author[b,c]{Sanjib Muhuri,}
\author[b]{Rama N. Singaraju,}
\author[b]{Jogender Saini,}
\author[b,d,e]{Tapan K Nayak,}

\affiliation[a]{Bhabha Atomic Research Centre, Mumbai -- 400085, India}
\affiliation[b]{Variable Energy Cyclotron Centre, Kolkata -- 700097, India}
\affiliation[c]{Homi Bhabha National Institute, Mumbai -- 400094, India}
\affiliation[d]{National Institute of Science Education and Research, Jatni -- 752050, India}
\affiliation[e]{CERN, Geneva 23, Switzerland}

\emailAdd{souravm@barc.gov.in}

\abstract{We present the research and development work of the first version of a 6$\times$6 array of silicon pad detectors, carried out in India, for the proposed forward calorimeter (FOCAL) as part of the ALICE collaboration upgrade program at CERN. The primary motivation is to develop a large area silicon pad array realizing the challenging requirements of high-energy physics experiments such as low leakage current, high breakdown voltage and evaluate its performance as an active layer in the prototype silicon tungsten (Si-W) electromagnetic (EM) calorimeter. Towards these goals, a 36-pad silicon sensor with an individual pad size of $\sim$1 cm$^2$ is designed on a 4-inch high resistivity N-type wafer and fabricated at Bharat Electronics Limited, Bangalore. The sensors have been used to assemble the prototype Si-W calorimeters and were successfully tested with high-energy particle beams. The design and development of the large area silicon sensor and its characterization using radioactive sources in the laboratory and high-energy particle beams are reported in this paper.}

\keywords{silicon pad array, pad sensor for calorimeter, large area silicon sensor}

\begin{document}
\maketitle
\flushbottom

\section{Introduction}
\label{sec:intro}
Silicon detectors, owing to their good energy and position resolution properties along with their insensitivity towards the magnetic field, are an attractive choice for the active detector layers, especially in vertex and tracking detectors and in calorimeters of a typical high-energy physics experiment with the presence of high magnetic field~\cite {silicon-1,silicon-2,silicon-3}. However, the design of a large area silicon sensor for high-energy physics experiments involves technological challenges and demands specific requirements like high breakdown voltage, low leakage current etc., to be satisfied to alleviate the performance degradations due to long-term operation under heavy particle fluence.

Large breakdown voltage is essential to safely increase the operating voltage of the detector gradually over the years, to mitigate the phenomena like charge collection efficiency degradation, type inversion, increase of full depletion voltage~\cite {si-hep1,si-hep2,si-hep3}. However, for the detectors fabricated in a planar process~\cite {si-hep4}, the breakdown voltage is limited by the presence of a high electric field at the junction edge~\cite {si-hep5,si-hep6} due to its sharp curvature. This requires additional design techniques to reduce the electric field crowding. Similarly, a low leakage current at the operating voltage is crucial to reduce the shot noise contribution from the detector, improving the signal to noise ratio (SNR) and facilitate better detection of the minimum ionizing particle (MIP). But inherent crystal lattice defects, trap centers, backside charge injection, defect propagation during various fabrication steps make it very challenging to keep the leakage current below few tens of nA per cm$^2$. Hence calls for additional processing steps.

The ALICE Collaboration at the Large Hadron Collider, CERN has proposed a new silicon-tungsten~\cite {si-w1,si-w2,si-w3} electromagnetic calorimeter, called FOCAL ~\cite {FOCAL-loi}, as a part of the upgrade program. In the calorimeter, tungsten as a passive absorber initiates the EM shower, which subsequently progresses along with the depth of the calorimeter with fractional ($\sim 2\%$) energy absorption in the active silicon layers. The primary objective of the FOCAL detector is to provide optimum position and energy resolution of the photons and separate $\gamma$ from $\pi^0$ up to 200 GeV~\cite {we-1,we-2} of incident energy for the incoming particles. FOCAL consists of two different types of silicon detector layers, high granularity layers (HGL) and low granularity layers (LGL). For the LGLs, a large area ($\sim$ 40 cm$^2$) silicon pad sensor with an individual pad size of $\sim$ 1 cm$^2$ are adopted, and to extract the accurate position information, a few HGLs consisting of monolithic active pixel sensors (MAPS) with pixel size 30$\times$30 $\mu$m$^2$ are planned in the prototype calorimeter~\cite {maps}. Active silicon pad layers furnish the energy information when summed up layer-wise, thereby constructing complete shower properties. 

Various R\&D activities towards a similar kind of silicon pad sensor development as the active detection medium in the calorimeters are reported in the literature. One of them is by the CMS collaboration for the HGCAL upgrade ~\cite {hgcal}, where prototypes of hexagonal P-type pad sensors with different active thicknesses and sizes are developed on both 6-inch and 8-inch wafers ~\cite {brondolin, pree}. It is reported that with the hexagonal shape of silicon pad development on a P-type starting wafer, a reduction in the overall dead area and an improvement in terms of radiation hardness has been achieved. On the other hand, the CALICE collaboration ~\cite {calice} has used silicon pad sensors of 6x6 cm$^2$ area with 1 cm$^2$ of individual pad size, developed on 4-inch wafers, in the physics prototypes for the Si-W EM calorimeter to be operated at a future linear electron positron collider ~\cite {daniel}. In the subsequent technological prototypes, they have used silicon pad sensors of 9$\times$9 cm$^2$ area with 5$\times$5 mm$^2$ of individual pad size developed on 6-inch high resistivity wafers ~\cite {jeremy}. An optimized guard ring structure has been reported in this work to reduce the dead area at the sensor edge and minimize the guard ring-peripheral pad cross talk.

This paper describes the development of a large area 36-pad silicon sensor intended for the low granularity layers of the FOCAL detector along with the test and characterization results.
Section 2 presents the design of the silicon pad array. Section 3 briefly reports the TCAD simulation study of a miniature silicon pad. The fabrication and packaging of the sensors are described in section 4. In section 5, testing and characterization of the fabricated sensors are discussed in detail. Finally, section 6 concludes the paper with a summary.

\section{Design of silicon pad array}
\label{sec:design}
\begin{figure}[htbp]
\centering 
\includegraphics[width=1.0\textwidth]{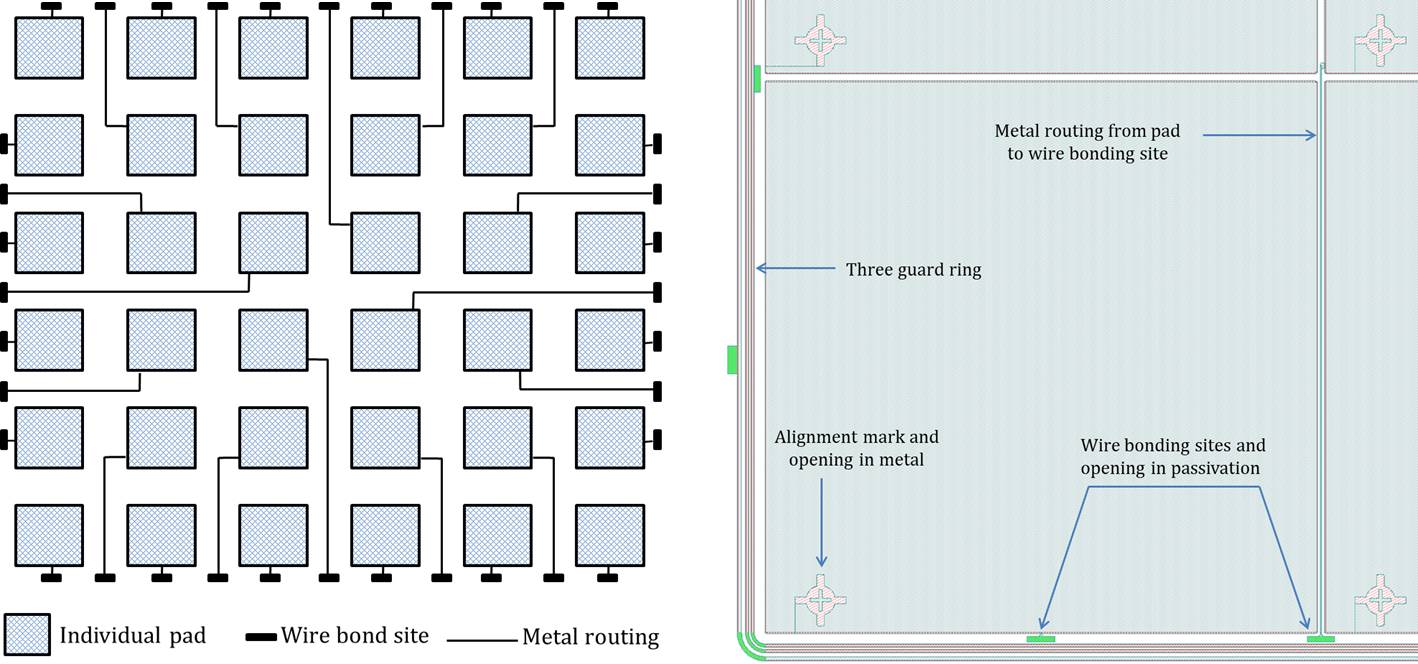}
\caption{\label{figure-1}(left panel) Routing scheme in a single metal from individual pad to wire bond sites (Distances are not to scale), (right panel) Detailed corner view of the 6x6 silicon pad array.}
\end{figure}

This design of the silicon pad sensor consists of four mask sets, i.e., P+, contact, metal and passivation-layer-opening, which are used to define the P+ regions, contacts, metallization and openings in the passivation layer. It is a "P on N" type design where the starting wafer is N-type, and a P-type dopant is implanted from the top to form the P-N junction. All the metal routing tracks from the individual pad to the wire bond sites at the periphery have been carried out in a single metal layer (due to foundry limitation), avoiding overlap, as shown in the left panel of figure~\ref {figure-1}.
The development of silicon pad array has been carried out with the flexibility of two different packaging options, first with conventional wire bonding and the second with flip-chip bonding using UBM (Under Ball Metallization)~\cite {ubm}. The metal and passivation-layer-opening mask sets are different for these two different designs. Development of silicon pad sensor with wire bond packaging option is reported in this paper. 

\begin{figure}[htbp]
\centering 
\includegraphics[width=1.0\textwidth]{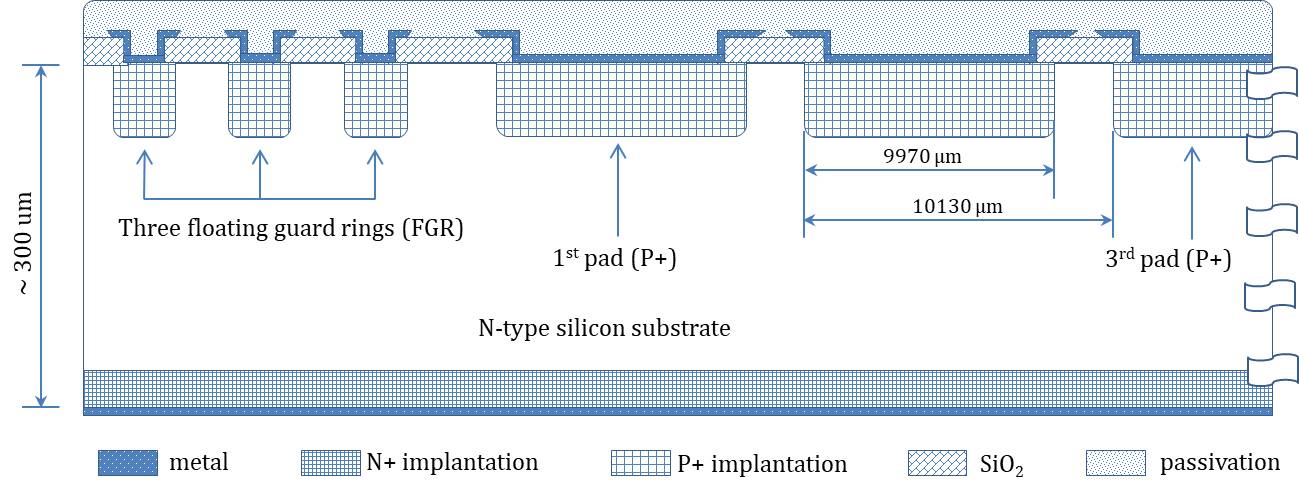}
\caption{\label{cross} Cross-section of the silicon pad sensor: three pads with the peripheral region (Distances are not to scale).}
\end{figure}

The total size of the sensor is $\sim$ 63 mm $\times$ 63 mm, where the actual size of each pad (P+) is 0.997 cm $\times$ 0.997 cm (optimized according to the study using Geant4 geometry simulation~\cite {radhe-syam}). The shape of the individual pad is a square type with rounded corners to avoid premature edge breakdown (PEB). Each pad consists of an alignment mark with an opening (window) in metal to check the integrity of individual pad connections. Windows in the metal layer allow tests with a laser by observing the photo-response. The right panel of figure~\ref {figure-1} shows one corner view of the sensor, and the corresponding cross-section of the 6$\times$6 silicon pad array is shown in figure~\ref {cross}. The sensors are passivated (0.45 $\mu$m), with openings in the wire bonding pad sites for testing (electrical characterization like I-V and C-V) and wire bonding. The passivation layer provides electrical stability to the sensor.

Three different design techniques, such as floating guard rings (FGR), metal overhang (MO) and deep junction depth~\cite {fgrmo-1,fgrmo-2,fgrmo-3,fgrmo-4,fgrmo-5,fgrmo-6,fgrmo-7}, are implemented to ensure sufficiently high breakdown voltage. FGR redistributes the potential around the junction over a wide area covered by the guard rings, while the MO approach distributes the electric field and minimizes the corner effects. In addition, with deeper junction depth, larger is the radius of curvature (spherical) of the junction at the edge, leading to reduced electric field crowding. 

This pad sensor is designed with three FGRs along the periphery surrounding the 6$\times$6 pads, 10 $\mu$m of MO over the individual P+ pads, and junction depth of 3 $\mu$m to have a high breakdown voltage. As shown in the right panel of figure~\ref {figure-1}, the option for biasing the FGRs is also accommodated (wire bond pads) in the design. Furthermore, additional fabrication steps like intrinsic and extrinsic gettering and back-plane ohmic side processing by double implantation are incorporated to achieve low leakage current~\cite {fgrmo-2,gettering}.

\section{TCAD simulation}
\label{sec:tcad}
\begin{figure}[htbp]
\centering 
\includegraphics[width=1.0\textwidth]{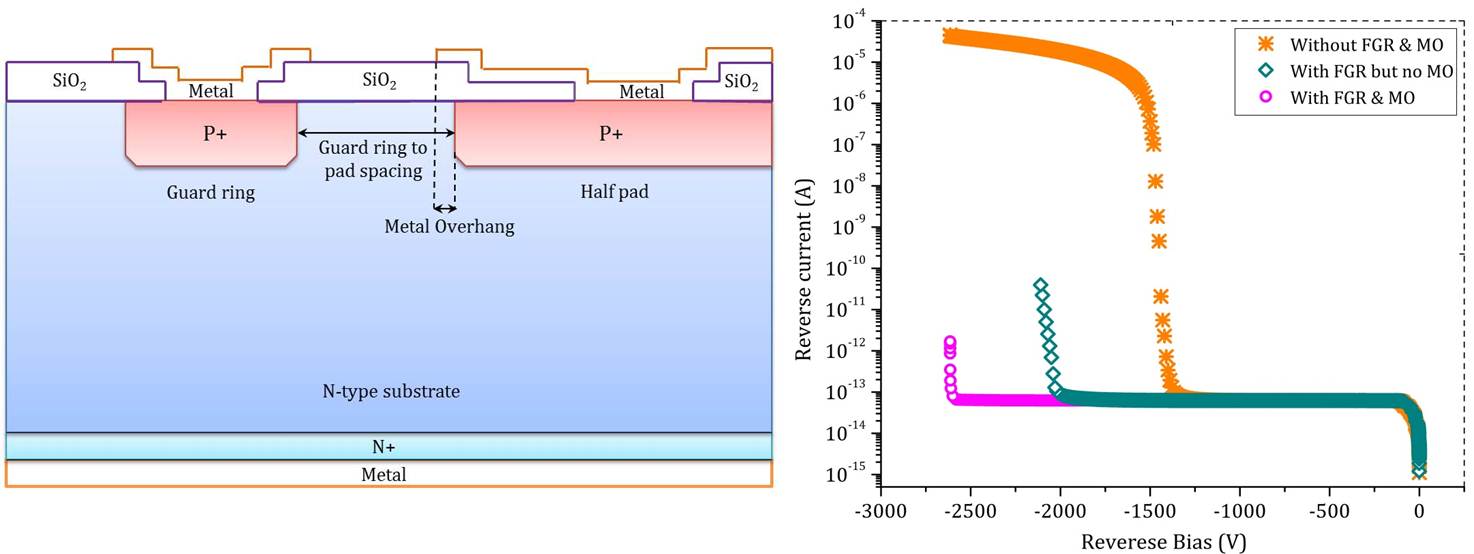}
\caption{\label{tcad-simulation} (left panel) Miniature simulation structure with half a pad and one guard ring, (right panel) Reverse I-V characteristics of the simulated pad structure showing improvements of breakdown voltage with the incorporation of design elements like FGR and MO: (orange) bare pad without FGR and MO, (green) pad with FGR but no MO, (pink) pad with both MO and FGR.}
\end{figure}

A miniature model (due to the maximum number of grid points limitation) of a single pad has been constructed through process simulation in a TCAD tool. Initially, a 2-D simulation of a single pad was performed and then subsequently, additional design elements like FGR and MO were introduced in the simulation. Process parameters (thermal cycles, implantation dose, and energy) w.r.t different fabrication steps were tuned, followed by the device simulation to analyze the electrical characteristics. Accordingly, geometrical design features (junction depth, guard ring width, spacing, and metal overhang) were determined. The left panel of figure~\ref {tcad-simulation} shows the cross-section of the simulated miniature pad structure with FGR and MO. One half of a pad (half-pad) along with a guard ring was simulated. The simulated structure has been optimized with the following geometrical parameters as tabulated in table~\ref {tab:1}.

\begin{table}[htb]
\centering
\footnotesize
\caption{\label{tab:1}Design parameters optimized during simulation}
\begin{tabular}{c c }
\hline \hline
  Parameters & Values \\ \hline  
  Half-pad width	& 60 um    \\ 
  Guard ring width & 40 um  \\ 
  Pad to guard ring spacing & 40 um  \\ 
  Metal overhang & 10 um  \\   \hline
\end{tabular}
\end{table}

The successive improvement in breakdown voltage with the introduction of FGR and MO is plotted in the right panel of figure~\ref{tcad-simulation}. A maximum breakdown voltage of $\sim$ 2.5 kV is achieved with the design incorporating both the FGR and MO, whereas the bare pad without FGR and MO has a breakdown voltage of $\sim$ 1.3 kV. The design incorporating only FGR has a breakdown voltage of $\sim$ 2 kV. 

\section{Fabrication and packaging}
\label{sec:packaged}
\begin{figure}[htbp]
\centering 
\includegraphics[width=0.85\textwidth]{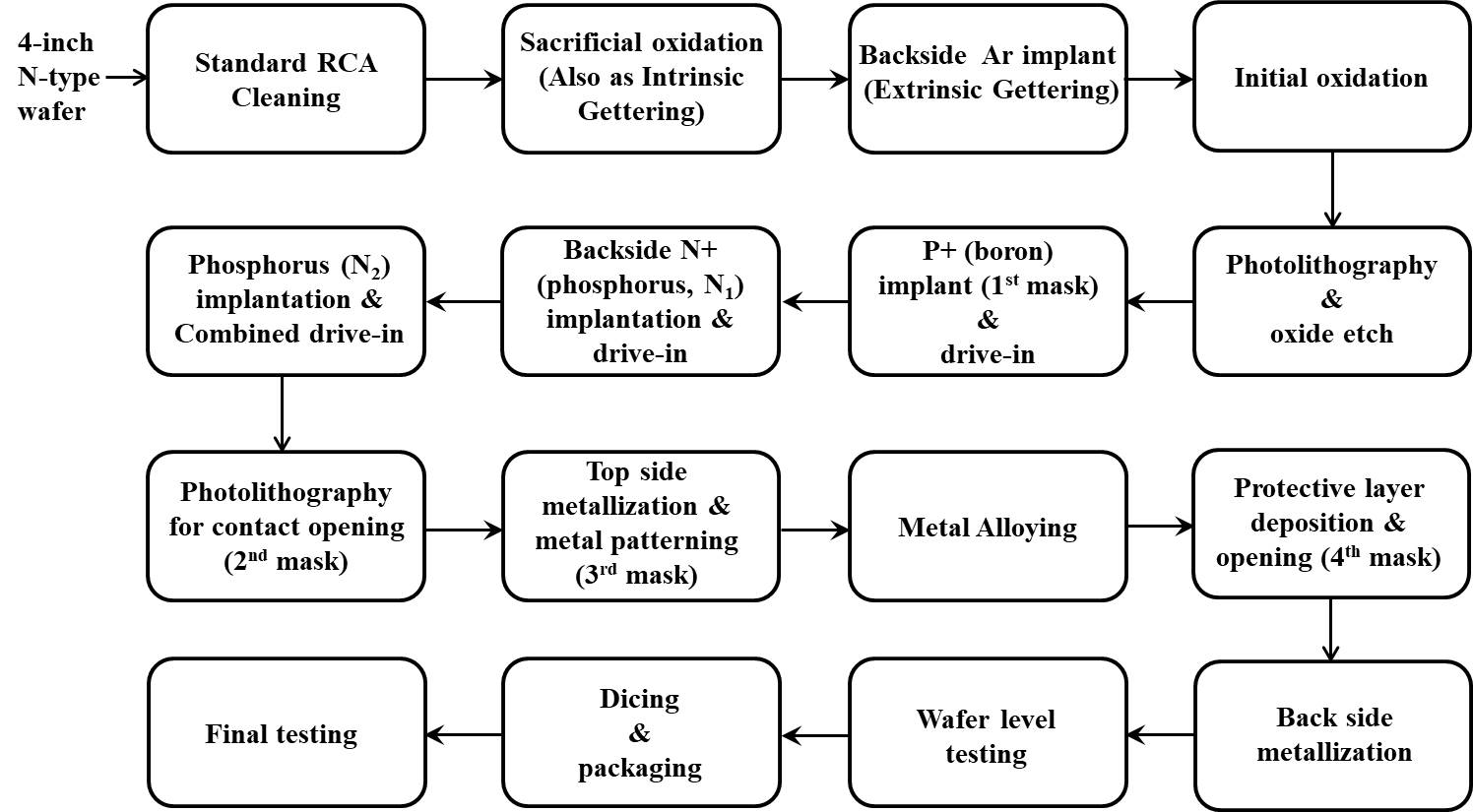}
\caption{\label{process} Process flow chart followed during the fabrication of 6$\times$6 array of silicon pad at the foundry. }
\end{figure}

The silicon pad sensors were fabricated on 4 inch N-type Float-Zone (FZ) single side polished wafers with $\langle 111 \rangle$ orientation, 3–5 k ohm-cm resistivity, and 300 ± 20 $\mu$m thickness. A flow chart of the typical sequence of processes followed during the detector development cycle is shown in figure~\ref{process}. The fabrication process starts with a standard wafer cleaning procedure followed by sacrificial oxidation (also acts as an intrinsic gettering thermal cycle) and etching out to remove the surface impurity. Argon implantation (extrinsic gettering) with high energy and high dose is then performed from the backside to purposefully create a damaged region that acts as a gettering site. Subsequently, initial oxidation, boron implantation, and drive-in steps have been carried out using the first mask (P+), followed by back-plane ohmic side processing by the double implantation method and combined drive-in steps. Then contact opening using the second mask (contact) and metal deposition and patterning using the third mask (metal) are performed. Finally, a passivation layer is deposited, and openings in the passivation layer in the bond pads have been created with the fourth mask (passivation-layer-opening).

\begin{figure}[htbp]
\centering 
\includegraphics[width=0.45\textwidth]{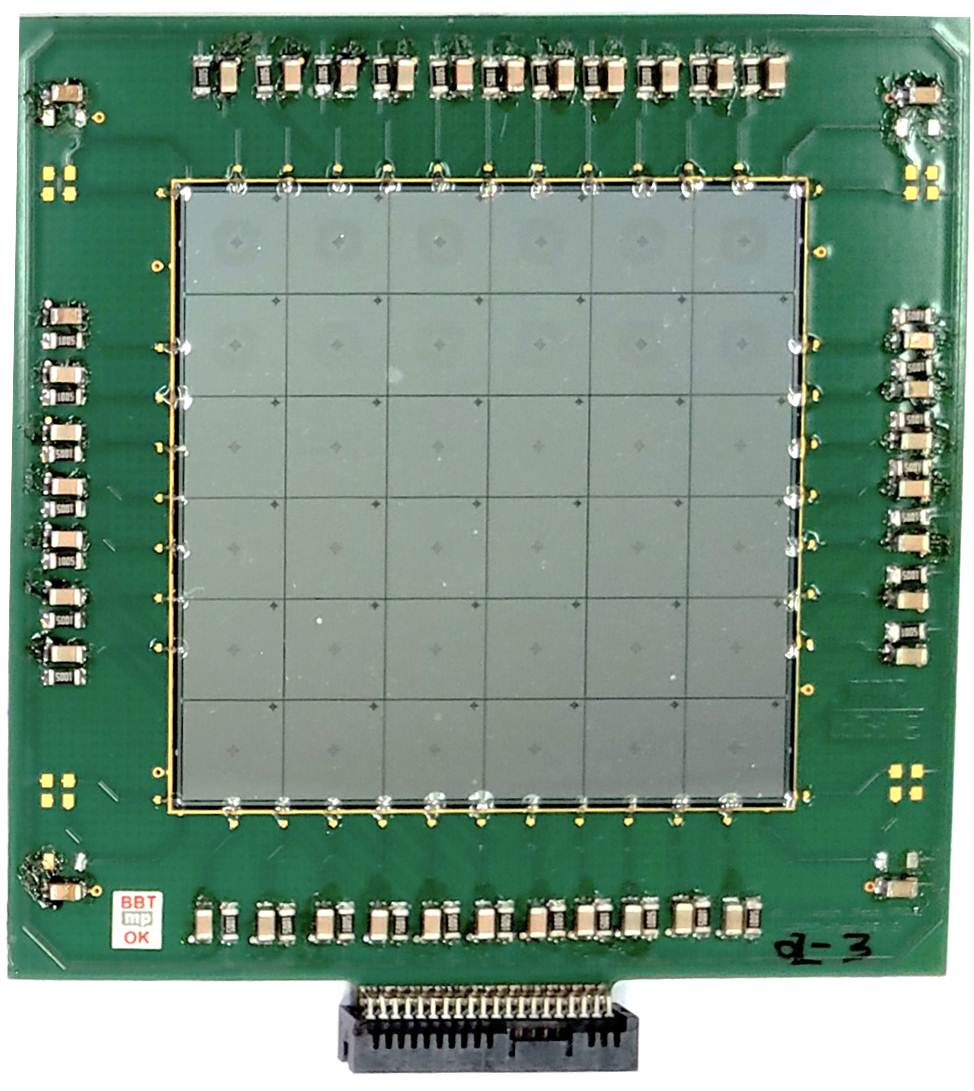}
\caption{\label{packaged_detector} Photograph of a fabricated and packaged silicon pad sensor mounted on detector PCB to be used for the prototype FOCAL.}
\end{figure}

During the packaging process, the 6$\times$6 silicon pad array was first diced from the fabricated wafer. The diced silicon pad sensor then underwent the process of "die-attach" on a thin PCB (0.8 mm) with a silver-filled, highly conductive (resistivity 0.006 ohm-cm) epoxy and then kept in an oven for over-night curing. Gold wire bonding from each wire bond site on the sensor to the bond pad on the PCB was then carried out, followed by applying glue (non-conductive epoxy) for protection. Care has been taken throughout the packaging process, during manual handling, and in choice of materials such as glue to ensure that the leakage current of the packaged sensor does not degrade much from its wafer-level value and is within the targeted limit of few tens of nA/cm$^2$. Figure~\ref{packaged_detector} shows the picture of the fabricated sensor, die attached, and wire bonded on a PCB (detector PCB). High Voltage (HV) bias lines for the pads and all the AC coupled signals from the sensor are distributed through a 40 pin connector to a back-plane PCB (as explained in section~\ref{sec:testing}).

\section{Characterization of sensors}
\label{sec:testing}
Testing of the 36-pad sensor was performed in three steps. Initially, electrical characterization was carried out at the wafer level to find out the operating parameters of the pads, i.e., breakdown voltage (V$_{BD}$), leakage current (I$_L$), full depletion voltage (V$_{FD}$), and junction capacitance at V$_{FD}$. Subsequently, laboratory tests were done with a radioactive source to determine the optimum operating voltage for the pads. Finally, the silicon pad sensors along with the tungsten layers are assembled to build a prototype FOCAL to carry out the sensor characterization by studying response from low energy deposition to high-energy deposition, i.e., response with minimum ionizing particle (MIP) and electron beam of a wide range of incident energies using the SPS beamline facility at CERN.

\subsection{Electrical characterization}
During electrical characterization, current versus reverse voltage (I-V) and detector junction capacitance versus reverse voltage (C-V) measurements of all the individual pads of multiple silicon pad sensors have been carried out. A semiconductor parameter analyzer (mainframe model 4200A-SCS) of Keithley/Tektronix was used for I-V /C-V  measurement. The leakage current was measured using an additional Keithley/Tektronix 2410 High power source measure unit (SMU), which in combination with 4200 can source and measure current from 10 fA to 1 A and voltage from 100 nV to 1100 V. In addition, Keithley/Tektronix 4210 high power SMU and CVU (Capacitance Voltage measurement Unit) were used during C-V measurement.

\begin{figure}[htbp]
\centering 
\includegraphics[width=1.0\textwidth]{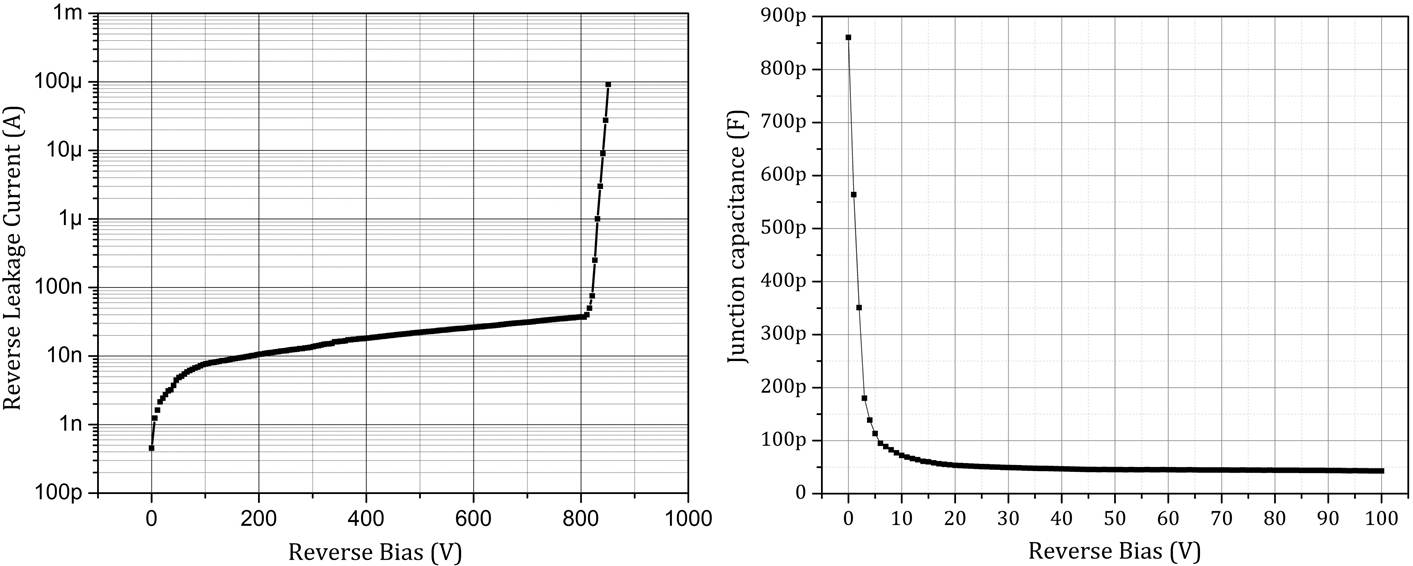}
\caption{\label{iv-cv}(left panel) Reverse I-V plot and (right panel) C-V curve of a particular pad among the 6x6 array of a wafer.}
\end{figure}

The left and right panels of figure~\ref{iv-cv} show the reverse I-V and C-V characteristics of a particular pad, among the 36 pads measured on a processed wafer. With full depletion achieved around 45-50 V, the pad shows $\sim$ 45 pF full depletion capacitance. It can be observed that the pad has less than 10 nA/cm$^2$ leakage current at full depletion voltage with a breakdown voltage of more than 800 V. This value of the breakdown voltage in the fabricated pad was less than the value achieved in the TCAD simulation. It may be due to the miniature simulation structure, simplified simulation without advanced/accurate defects density/trap centre model, or a possible mismatch between the simulation and the actual fabrication process environment. However, the achieved results (V$_{BD}$ of more than 800 V and less than 10 nA/cm$^2$ of I$_L$) are satisfactory, meeting the required specifications.

\begin{figure}[t]
\centering 
\includegraphics[width=0.75\textwidth]{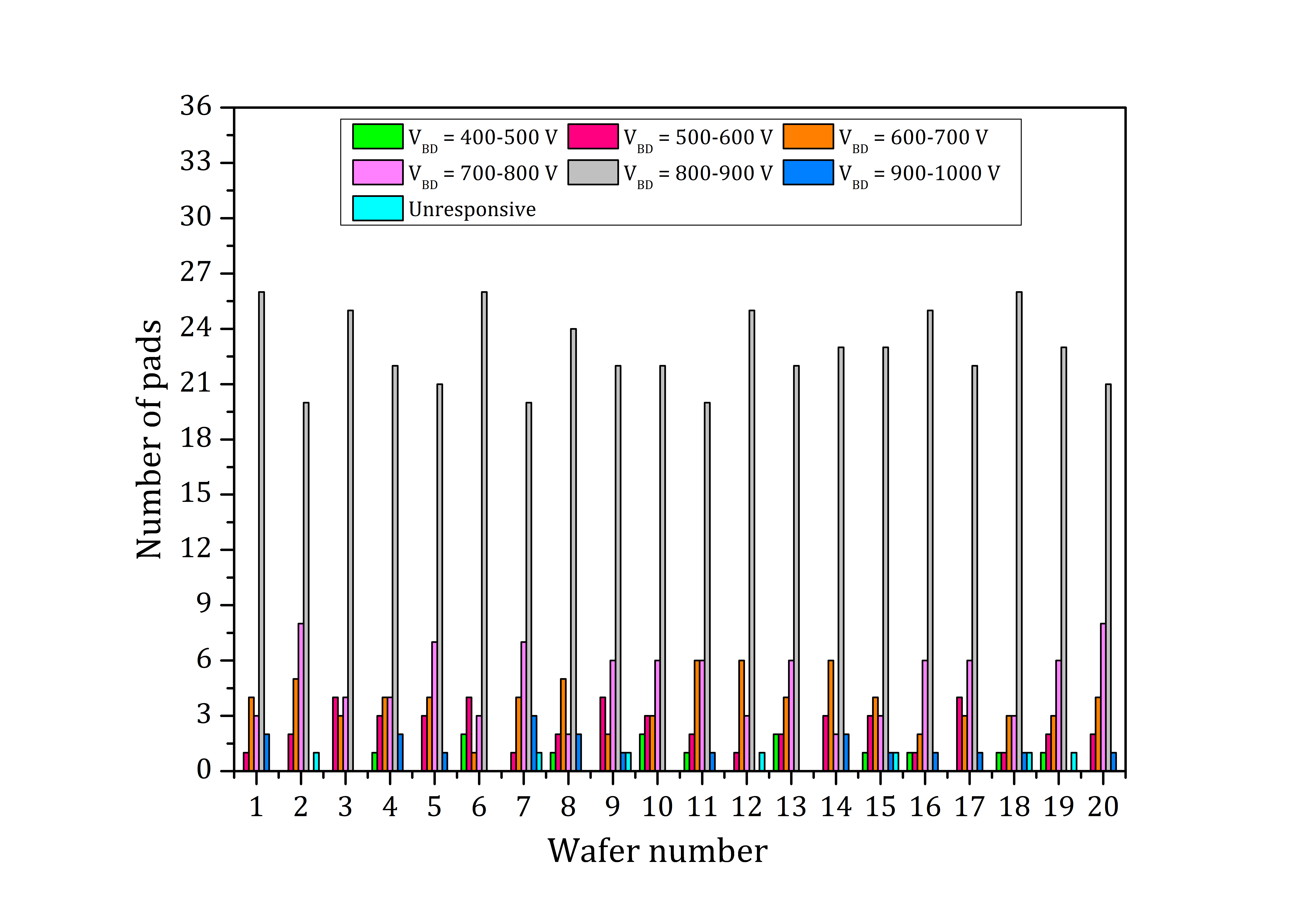}
\caption{\label{break-uni} Uniformity plot for the breakdown voltage showing the spread among the 6$\times$6 pads and across the wafers; significantly in all these 20 wafers, at least $\sim 95\%$ of the pads in a single wafer are having a breakdown voltage more than 500 V while most of them are having a breakdown voltage in the range of 800-900 V.}
\end{figure}

\begin{figure}[b]
\centering 
\includegraphics[width=0.75\textwidth]{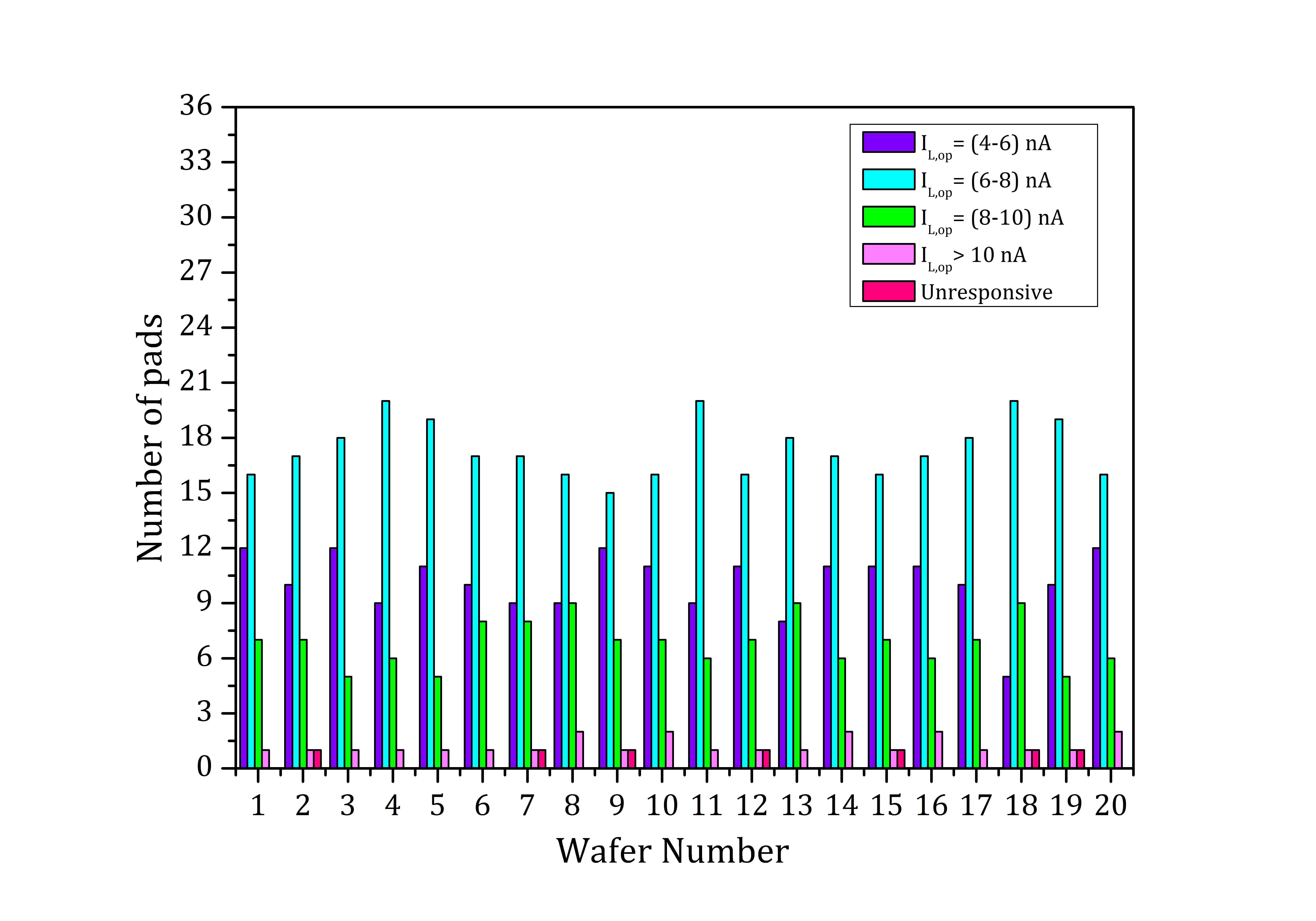}
\caption{\label{leak-uni} Uniformity plot for the leakage current showing the variation among the 6$\times$6 pads and across the wafers. At least $\sim 95\%$ of the pads in a single wafer are having a leakage current of less than 10 nA at the operating voltage.}
\end{figure}

Likewise, the I-V and C-V measurements were performed on all 36 pads of multiple wafers. The breakdown voltage (V$_{BD}$) and the leakage current of the pads at an operating voltage (I$_{L,op}$) of 15 V above the full depletion voltage (V$_{FD}$) were observed to determine the characteristic uniformity among the 36 pads and across the wafers. The acceptable wafers were chosen with the criterion of not more than two pads (including the unresponsive ones) in a wafer not conforming to the targeted specifications of V$_{BD}$ of more than 500 V and I$_{L,op}$ of less than 10 nA/cm$^2$. The results of twenty such acceptable wafers, which were later used to build the prototype FOCAL, are plotted in figure~\ref{break-uni} and figure~\ref{leak-uni}. The consistent nature of detector performance and characteristic uniformity across the wafers, as evident from these figures, signifies good detector design and proper processing with tight parameter control and careful handling during detector fabrication and packaging. 

Around 50 wafers have been processed during this large area pad sensor development. The acceptable wafers were selected based on the criterion mentioned above. Among them, few wafers later needed to be kept aside during and after packaging. Finally, 20 packaged silicon pad sensors met the criterion and were utilized to build the prototype FOCAL detector. Thus, out of the 50 wafers processed in the fab, the production yield factor is about $\sim40\%$.

The salient physical and technical features of the fabricated and tested pad sensor are listed in table~\ref {tab:2}.

\begin{table}[htbp]
\centering
\footnotesize
\caption{\label{tab:2}Physical and technical specifications of the silicon pad sensor}
\begin{tabular}{c c }
\hline \hline
Parameters & Values \\ \hline  
Wafer size	 &  4-inch    \\ 
No of pads & 36\\
Wafer thickness &	 300 $\mu$m+/- 20 $\mu$m \\
Wafer resistivity	&  3-5 kohm-cm \\
Crystal orientation  &	$\langle 111 \rangle$  \\
Total area	&  62.65 cm x 62.65 cm \\
Polishing	& Single-sided \\
Leakage current at operating voltage	& $<$ 10 nA/cm$^2$ \\
Breakdown voltage &	$>$ 500 V \\
Full depletion voltage (V$_{FD}$) &	45-50 V \\
Junction capacitance at V$_{FD}$ &	$\sim$ 45 pF \\ \hline
\end{tabular}
\end{table}

\subsection{Test with beta source and high-energy beam}
\begin{figure}[htbp]
\centering 
\includegraphics[width=0.95\textwidth]{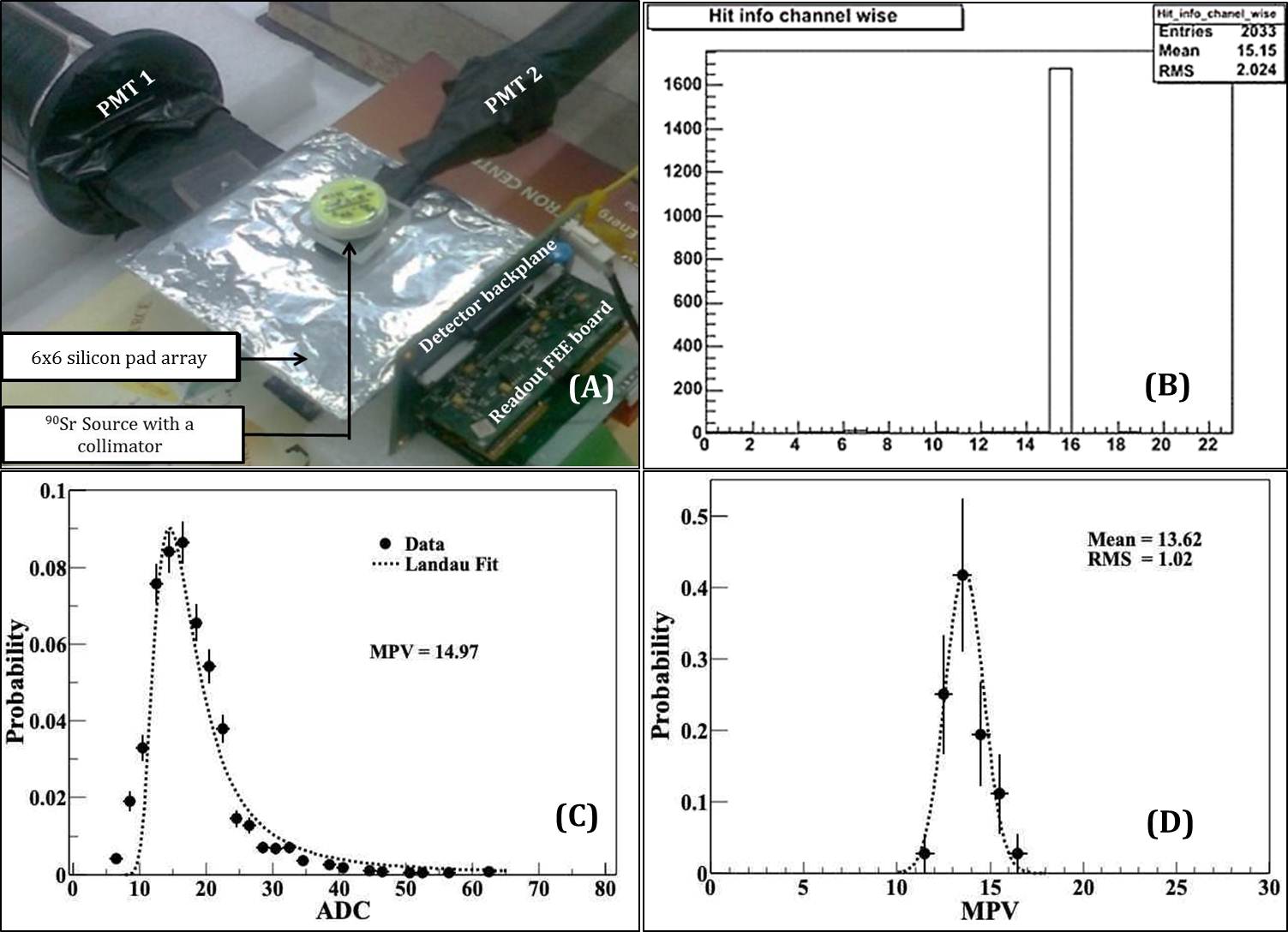}
\caption{\label{9} (A) Laboratory test setup for the silicon pad sensor characterization. (B) Single pad hit, mapping. (C) A typical response of a silicon pad fitted with a landau distribution. (D) Distribution of the most probable value of energy loss in different pads.}
\end{figure}

In the next step, the pad sensors were characterized with low energy electrons using a $^{90}$Sr radioactive source (with endpoint energy of 0.546 MeV) at the laboratory. An experimental setup consisting of two-fold coincidence trigger scintillators, one finger scintillator, detector PCB and front end electronics (FEE) readout boards, as shown in figure~\ref{9}A, was used to characterize the pads at the laboratory. A packaged sensor is first wrapped in a black sheet and then covered with Aluminium foil for shielding against EMI, and was coupled with the readout electronics via a back-plane PCB which was designed specifically to assemble four sensor modules and FEE boards containing ASIC MANAS ~\cite {manas} and ASIC ANUSANSKAR ~\cite {we-1}. The pad signals, read out by the FEE boards, were sent to the PCICFD data acquisition system~\cite {pcicfd} through a translator board (translates the LVTTL signal from FEE to LVDS signal and LVDS signal back to LVTTL). As shown in figure~\ref{9}B, the result represents the number of hits as a function of pad index (channel number). The peak around pad index 16 indicates the position of the source. Figure~\ref{9}C shows the typical response of the targeted silicon pad (pad index 16) with $^{90}$Sr. Fitted with a landau distribution, the most probable value (MPV) of the energy loss is 14.97 in units of ADC count (1 ADC count corresponds to $\sim$ 0.6 mV; a 12 bit ADC AD7476 with 2.5 V reference voltage). Likewise, the responses of various pads were tested by shifting the source position along with the scintillators manually. Faulty pads were identified and marked for future reference. The responses of the pads found to be normalized, as shown in figure~\ref{9}D, during the test with beta source. It represents the distribution of the MPV of energy loss in different pads with a mean of 13.62 and RMS of 1.02 in units of ADC count. This figure emphasizes the response uniformity among the pads.

\begin{figure}[htbp]
\centering 
\includegraphics[width=0.9\textwidth]{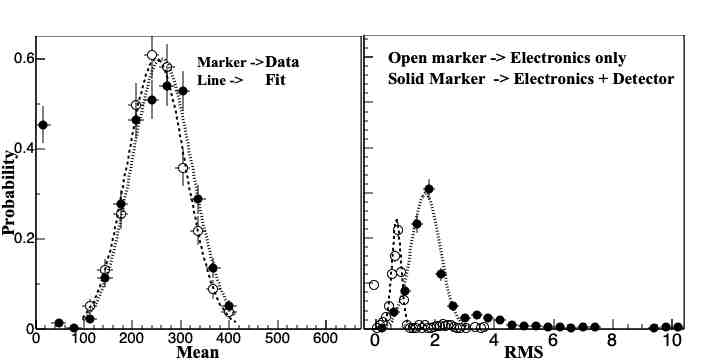}
\caption{\label{ped} The distribution of the mean (left panel) and RMS (right panel) values of the pedestals, measured during the stand-alone test of the FEE chain and with the biased pad sensor connected to it \cite {we-2}.}
\end{figure}

Figure~\ref{ped} shows the distribution of the mean and RMS of the pedestals measured during a stand-alone test of the FEE chain and with biased pad sensors connected to it. The RMS value of the pedestals, measured during the stand-alone test of the front-end electronics, is typically around 0.8 ADC counts. It increases to $\sim$ 2.0 ADC counts when the biased silicon pad sensors are connected to the FEE input. This plot assures minimal contribution from the FEE chain in the RMS value of figure~\ref{9}D.

\begin{figure}[htbp]
\centering 
\includegraphics[width=0.95\textwidth]{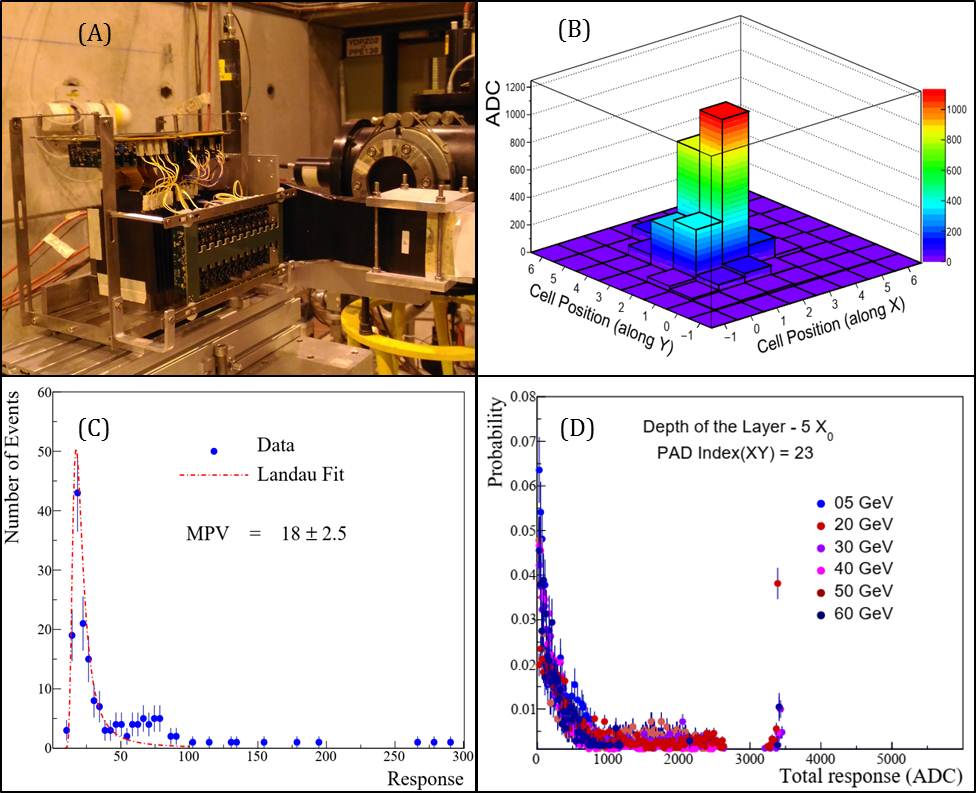}
\caption{\label{13}(A) Experimental setup at SPS-CERN. (B) 3-D plot showing the energy deposition in terms of ADC count for different pads in a particular sensor layer (layer eight among the 19 layers) in response to the electron beam of 30 GeV incident energy. (C) The Typical response of a pad in a particular sensor layer to 120 GeV pion beam (MIP) \cite {we-2}. (D) Response of pad 23 (array coordinate 2,3) of the sensor situated in the 8th layer for a wide range of incident energies of the incoming electron beam.}
\end{figure}

A Prototype Si-W sampling type electromagnetic calorimeter was built with 19 layers of alternating 6x6 array of silicon pad detectors and tungsten absorber/converter as a very next step \cite {we-2}. As shown in figure~\ref{13}A, the experimental setup was installed at the H6 beamline facility at CERN-SPS~\cite {sps}. The sensors in the prototype were read out using MANAS and newly developed ANUSANSKAR ASICs. Both the ASICs are multi-channel pulse processors with multiplexed output. The ANUSANSKAR ASIC has an enhanced dynamic range of +/- 600 fC~\cite {we-1} compared to -300 fC to +500 fC for existing MANAS. The setup used a two-fold coincidence signal as the trigger unit. A finger scintillator was used to pinpoint the collimated beam to the detector. The sensors were exposed to different (electron, pion) beam types for a wide range (20 - 60 GeV, 120 GeV) of incident energies. It helps to understand the silicon pad response from very low energy (pion) to large energy deposition. The electromagnetic shower, developed within the calorimeter, might result in very high-energy deposition even within a single pad. Both the extreme cases with pion and EM-shower are essential to be sensed by the silicon pads, which might affect the calorimeter responses. The response of a pad to the high-energy pion beam, known as minimum ionizing particle (MIP) response, illustrates its ability to filter out the minimum signal from the background noise.

Figure~\ref{13}C shows the MIP response of a single pad for a particular sensor layer. As expected, the MIP signal can be understood in terms of a landau distribution and resulted in an 18 ADC count as the most probable value for a thin detector. On the other hand, the response of the pad sensor to electrons entering into the calorimeter can be explained in terms of response to EM shower. Electrons, being lighter in mass and charged particles, initiates an EM shower, propagating along with the calorimeter depth and confined within a characteristic length (Molière radius) in the transverse direction. The energy deposited by the shower depends on the depth it traversed. It rises for few first layers and then follows a falling trend that leaves a peak with a value that varies as a function of the incident electron's energy. Unlike the test at the laboratory, more than one pad in a layer might get hits because of the transverse extension of the shower, which is presented in the two-dimensional hit distribution, weighted with ADC count as shown in figure~\ref{13}B. Figure~\ref{13}D represents the distribution of energy deposition, as a function of energy of the incoming electrons, for the pad (array coordinate 2,3) situated in the 8th layer (among the 19 layers) and lying at the centroid of the shower. 

\section{Summary $\&$ Conclusion}
\label{sec:conclusion}
The development of a large area ($\sim$ 40 cm$^2$) N-type 36-pad silicon sensor on a 4-inch high resistivity wafer is presented in this paper. The fabricated sensors show good uniformity w.r.t breakdown voltage and leakage current among the pads and across the wafers, with a production yield factor of $\sim$ 40$\%$. A breakdown voltage of more than 500 V and a leakage current of less than 10 nA/cm$^2$ at an operating voltage of $\sim$ V$_{FD}$+15 V for most of the pads ($\ge$95$\%$) in a sensor has been observed. The packaged sensors were later tested in the SPS beamline at CERN. A prototype electromagnetic calorimeter, using these 6$\times$6 silicon pad detectors, was exposed to an electron beam up to an energy of 60 GeV, and the calorimetric performance was found to be satisfactory. The pad sensors were also exposed to a pion beam and produced a MIP-like response. 

In this version of the silicon pad sensor, routing of long metal lines from the inner pads to the wire bonding sites at the periphery leads to an increase in pad-pad spacing and a reduction in the fill factor. The separation between the innermost guard ring and the pads in the periphery has to be modified to accommodate the wire bonding pads. In the next planned version of the large area silicon pad sensor on a 6-inch wafer, a through-hole flip-chip wire bonding~\cite {cms-pad} packaging scheme is envisaged to improve the fill factor and enhance the breakdown voltage. Inter pad guard ring (with additional metal layers) could be another design option for further improving the leakage current.

\acknowledgments
We want to thank Ranjay Laha, Arijit Das, Y P Prabhakar Rao and Rajeena Rani of Bharat Electronics Limited (BEL), Bengaluru, for fabrication support. We also thank all the crew members of the SPS beamline for the excellent quality of the beam for the detector test and ALICE-FOCAL collaboration for support during the tests.

\end{document}